\documentclass[apjl]{emulateapj}

\usepackage{apjfonts}

\shorttitle{Discovery of a bright T dwarf}
\shortauthors{Artigau et al.}

\begin{document}

\title{Discovery of the brightest T dwarf in the northern hemisphere}

\author{\'Etienne Artigau\altaffilmark{1,2}, Ren\'e Doyon\altaffilmark{1}, David Lafreni\`ere\altaffilmark{1}, Daniel Nadeau\altaffilmark{1}, \\ Jasmin Robert\altaffilmark{1} and Lo\"ic Albert\altaffilmark{3}}
\altaffiltext{1}{D\'epartement de Physique and Observatoire du Mont M\'egantic, Universit\'e de Montr\'eal, C.P. 6128, Succ. Centre-Ville, Montr\'eal, QC, H3C 3J7, Canada}
\altaffiltext{2}{Gemini Observatory, Southern Operations Center, Association of Universities for Research in Astronomy, Inc., Casilla 603, La Serena, Chile}
\altaffiltext{3}{Canada-France-Hawaii Telescope Corporation, 65-1238 Mamalahoa Highway, Kamuela, HI 96743}
\email{artigau@astro.umontreal.ca doyon@astro.umontreal.ca \\ david@astro.umontreal.ca nadeau@astro.umontreal.ca \\ jasmin@astro.umontreal.ca albert@cfht.hawaii.edu}

\begin{abstract}

We report the discovery of a bright ($H=12.77$) brown dwarf designated SIMP~J013656.5+093347. The discovery was made as part of a near-infrared proper motion survey, SIMP (Sondage Infrarouge de Mouvement Propre), which uses proper motion and near-infrared/optical photometry to identify brown dwarf candidates. A low resolution ($\lambda/\Delta\lambda\sim40$) spectrum of this brown dwarf covering the 0.88-2.35 $\mu$m wavelength interval is presented. Analysis of the spectrum indicates a spectral type of T$2.5\pm0.5$. A photometric distance of $6.4\pm0.3$ pc is estimated assuming it is a single object. Current observations rule out a binary of mass ratio $\sim$1 and separation $\gtrsim5$ AU. SIMP 0136 is the brightest T dwarf in the northern hemisphere and is surpassed only by $\varepsilon$~Indi~Bab over the whole sky. It is thus an excellent candidate for detailed studies and should become a benchmark object for the early-T spectral class.

\end{abstract}

%\noindent{\em Suggested running page header:} Discovery of a bright T dwarf

\section{Introduction}

Numerous L and T dwarfs have been found during the last decade through the large scale Two Micron All Sky Survey (2MASS, \citealp{Skrutskie2006}), the Deep Near-Infrared Survey of the Southern Sky (DENIS, \citealp{Epchtein1997}), and the Sloan Digital Sky Survey (SDSS, \citealp{York2000}). These objects have effective temperatures ranging from $\sim$2300 to $\sim$750 K and, with the exception of some early-L dwarfs, have substellar masses and are thus brown dwarfs (BDs). The L dwarfs are characterized by the presence of dust condensates suspended in their atmosphere and by increasingly red near-infrared (NIR) colors while the T dwarfs are distinguished by the appearance of broad methane absorption features in their NIR spectrum and by bluer NIR colors. The L/T transition is marked by a rapid decrease of $J-K_{\rm s}$, from $\sim$2 to $\sim$0 between late-L and mid-T, and by an increase of the $J$-band flux, in contrast with the overall decreasing trend from early-L to late-T \citep[see][and references therein]{Kirkpatrick2005}. Several scenarios have been proposed to explain these rapid variations of the NIR properties of BDs. \citet{Burgasser2002a} propose that they are due to the clearing of patches in the cloud decks, with increasingly clearer atmospheres for later types. \citet{Tsuji2003} suggest that, at the L/T transition, the $M_J$ versus $J-K_{\rm s}$ diagram does not constitute a single evolutionary sequence. Rather, it is populated by objects of different surface gravities, for which the sinking of dust clouds below the photosphere occurs at different values of $M_J$ and temperature. \citet{Knapp2004} offer an explanation based on a rapid rain-out of condensates from the photosphere occurring from late-L to mid-T. These scenarios make predictions that can be tested given a sufficiently large sample of early-T dwarfs.

Almost all known BDs were initially selected based on their NIR colors, and/or optical colors when available. Although this is an effective mean of identifying L or late-T dwarfs, early-T dwarfs are difficult to differentiate from M dwarfs when only NIR data are available because both types share similar $J-H$ and $H-K_s$ colors. Deep $i$- or $z$-band imaging, as in SDSS, removes the degeneracy since $i-z$ and $i-J$ differ significantly between M and T dwarfs. So far SDSS has covered 25\% of the sky and found eight\footnote{http://dwarfarchives.org/ \label{bdarchive}} T dwarfs of spectral type T4.5 or earlier and brighter than $J=16$ (a conservative completeness limit for 2MASS). On the other hand, BD searches based on the 2MASS point source catalog (2MASS PSC, \citealp{Cutri2003}) have unveiled only nine\footnotemark[\ref{bdarchive}] such objects over an area three times larger. This suggests that many early-T dwarfs remain to be indentified in the 2MASS PSC.

Proper motion (PM) is an efficient mean of discriminating BDs from M dwarfs. For a given magnitude, BDs are much closer and therefore have, on average, larger PMs. We have undertaken a NIR PM survey aimed at identifying new BDs and, in particular, increasing the sample of early-T dwarfs. This Letter reports the discovery, as part of this survey, of the brightest T dwarf in the northern hemisphere.

\section{The \textit{SIMP} near-infrared proper motion survey}

SIMP (Sondage Infrarouge de Mouvement Propre) is a proper motion survey made with the Observatoire du Mont M\'egantic (OMM) wide-field NIR camera CPAPIR (\'E. Artigau et al., in preparation) currently installed at the CTIO 1.5m telescope operated by the SMARTS consortium. The camera features a $35^{\prime}\times35^{\prime}$ field of view and a pixel scale of $1\farcs03$. The PMs of the sources detected are determined by comparing their measured position with the 2MASS PSC. The SIMP observations were initiated in 2005 February and are still ongoing. Since 2MASS started in 1997 June and ended in 2001 February, the time span between the two epochs ranges from $\sim$4 to $\sim$9 years. The 5$\sigma$ uncertainty on the relative SIMP and 2MASS astrometry is $1^{\prime\prime}$, equivalent to a PM lower limit of $0.125$-$0.25^{\prime\prime}$/yr, or a tangential velocity limit of 15-30~km/s at 25 pc. Up to the 2MASS $J$-band magnitude limit of $\sim$16.5, T-type dwarfs may be found out to $\sim$25 pc while L-type dwarfs may be located as far as 100 pc. Since solar neighborhood L and T dwarfs with known parallax have median tangential velocities of 25 km/s and 39 km/s, respectively \citep{Vrba2004}, almost all T dwarfs within 25 pc and roughly half of L dwarfs within 50 pc observed by SIMP should be recognized as such, provided that they are listed in the 2MASS PSC.

Full details of the SIMP survey will be presented in a future paper. At the time of writing, the survey has covered $\sim$4200 square degrees near the celestial equator. More than 300 BD candidates have been identified; $I$-band imaging and NIR spectroscopy is underway to confirm their spectral type.

\section{SIMP 0136}

\subsection{Discovery}

\begin{figure}
\epsscale{0.99}
\plotone{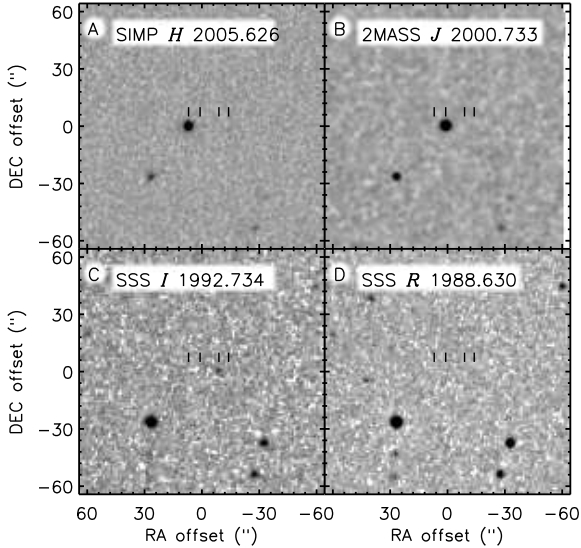}
\caption{\label{fig1} SIMP, 2MASS and SSS images of a $2^\prime\times2^\prime$ field centered on the J2000.0 position of SIMP 0136. In all panels, vertical lines are shown, from left to right, above the expected position of SIMP 0136 at the epoch of SIMP, 2MASS, SSS $I$ and SSS $R$ observations respectively, as derived from its measured PM.}
\end{figure}

\begin{figure}
\epsscale{.99}
\plotone{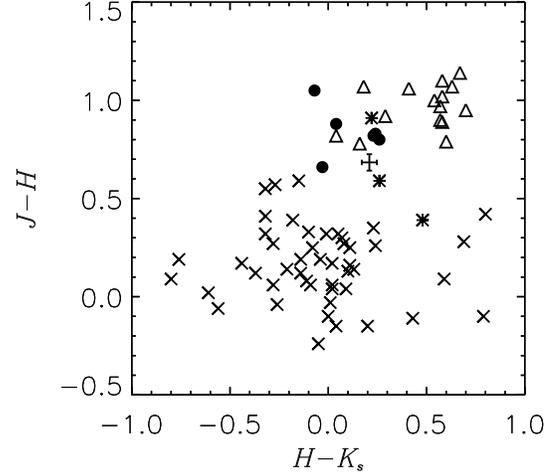}
\caption{\label{fig2} $J-H$ vs $H-K_s$ diagram of SIMP 0136 and all known T dwarfs with $J<16$. SIMP 0136 is identified by its error bars. Triangles, filled circles, stars and crosses respectively identify T0-T1.5, T2, T3-T3.5 and T4-T9 BDs.}
\end{figure}

Our initial analysis of the survey data quickly revealed a candidate with a very high PM and colors consistent with a T dwarf. The source SIMP~J013656.5+093347, hereafter abbreviated SIMP 0136, moved by $6\farcs3$ over the $\sim$5 years separating the 2MASS and SIMP observations. It is not seen in the SuperCOSMOS Sky Survey (SSS, \citealp{Hambly2001I}) $B$ or $R$ images, but it is detected in $I$ at a position consistent with the PM deduced from the 2MASS and SIMP measurements. Table~\ref{tbl-1} lists all the measurements available for SIMP 0136 from the 2MASS PSC, the SIMP image and the SSS catalog\footnote{http://www-wfau.roe.ac.uk/sss/}, as well as its calculated PM, spectral type (see Sect. \ref{spectro}), and photometric distance (see Sect. \ref{discussion}). Figure~\ref{fig1} shows SIMP, 2MASS and SSS images of a small field centered on SIMP 0136; the high PM of SIMP 0136 is clearly apparent. The SSS and 2MASS photometric measurements yield $I-J=5.07\pm0.17$, a value typical of early-T dwarfs. As shown in Fig.~\ref{fig2}, its NIR colors, $J-H = 0.68\pm0.04$ and $H-K_s = 0.21\pm0.04$, are also consistent with an early-T spectral type. 

\begin{table}
\begin{center}
\caption{\label{tbl-1} Properties of SIMP~J013656.5+093347}
\begin{tabular}{ll}
\hline\hline
Parameter&Value\\
\hline\hline
\multicolumn{2}{c}{SSS ($1988.630$)}\\
$R$&\textgreater$20.8^{\rm a}$ \\

\multicolumn{2}{c}{SSS ($1992.734$)}\\
$\alpha$, $\delta$&$01$:$36$:$55.962$, $+09$:$33$:$47.40$\\
$I$&18.52$\pm$0.17$^{\rm b}$\\

\multicolumn{2}{c}{2MASS PSC ($2000.733$)}\\
$\alpha$, $\delta$&$01$:$36$:$56.624$, $+09$:$33$:$47.32$\\
$J$&13.455$\pm$0.028 \\
$H$&12.771$\pm$0.031 \\
$K_s$&12.562$\pm$0.023 \\

\multicolumn{2}{c}{SIMP ($2005.626$)}\\
$\alpha$, $\delta$ & $01$:$36$:$57.046$, $+09$:$33$:$47.28$\\

\multicolumn{2}{c}{SIMP ($2006.470$)}\\
$\alpha$, $\delta$ & $01$:$36$:$57.115$, $+09$:$33$:$47.32$\\

\hline
$\alpha_{2000.0}$, $\delta_{2000.0}$  & $01$:$36$:$56.566$, $+09$:$33$:$47.30$\\
$\mu_{\alpha} \cos(\delta)$           & $+1241\pm9$~mas/yr\\
$\mu_{\delta}$                        & $-4\pm10$~mas/yr\\
Photometric distance & $6.4\pm0.3$ pc$^{\rm c}$\\
Spectral type        & T$2.5\pm0.5$\\
\hline
\end{tabular}
\tablenotetext{}{All coordinates are in J2000 equinox.}
\tablenotetext{a}{From \citet[Table~1]{Hambly2001I}.}
\tablenotetext{b}{Uncertainty from \citet[Table~12]{Hambly2001II}.}
\tablenotetext{c}{Assuming a single object.}
\end{center}
\end{table}

\pagebreak
\subsection{Spectroscopy}\label{spectro}

\begin{figure}
\epsscale{.99}
\plotone{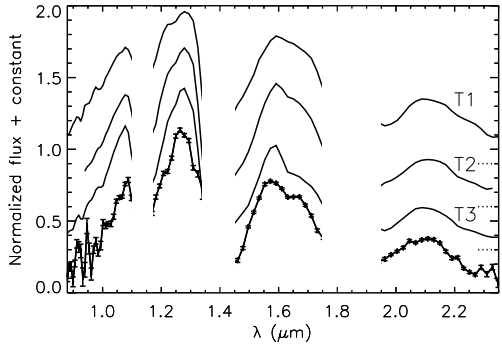}
\caption{\label{fig3} NIR spectrum of SIMP 0136 compared to the spectra of SDSS0151+12 (T1), SDSS1254-01 (T2) and 2MASS1209-10 (T3). Regions of the spectrum severely affected by telluric absorption have been masked out. Comparison spectra are from \citet{Burgasser2006} and \citet{McLean2003} and were binned to the same resolution as our observations. All spectra are normalized at 1.25 $\mu$m and offset by a constant.}
\end{figure}

We obtained NIR observations of SIMP 0136 on 2006 July 16 at the OMM 1.6-m telescope located in southern Qu\'ebec. The SIMON spectrograph (L. Albert et al., in preparation) was used with the Amici prism which provides simultaneous coverage of the  0.88-2.35$\mu$m wavelength interval at a $\lambda/\Delta\lambda$ ranging from 35 to 50. A total of 18 30-s exposures were obtained by dithering over five positions along the 1\farcs1 slit (2.4 pixels). Individual frames were flat-fielded using dome-flat images. The A0 star HD~9560 was observed to correct for telluric absorption. Observations of both SIMP 0136 and the spectroscopic standard were obtained at an average air mass of 1.45. Fig.~\ref{fig3} shows the resulting spectrum, which has a signal-to-noise ratio per spectral pixel ranging from 10 around 1~$\mu$m to more than 100 in the $H$-band.

The overall NIR spectral energy distribution (SED) of SIMP 0136 is similar to that of other early-T dwarfs (SDSS0151+12, SDSS1254-01, 2MASS1209-10, respectively T1, T2 and T3; \citealp{Geballe2002,Leggett2000,Burgasser2004}). The sharp SED fall-off below 1.1~$\mu$m due to the very large red wing of the K I doublet at 0.77~$\mu$m is evident despite the modest signal-to-noise. The $H$-band clearly shows CH$_4$ absorption between 1.6~$\mu$m and 1.7~$\mu$m, the hallmark of T dwarfs, with a strength intermediate between those of the comparison T2 and T3. The $K$-band flux is depressed beyond 2.2~$\mu$m due to CH$_4$ absorption.

Five spectral indices defined in \citet{Burgasser2006} can be calculated from our low resolution spectrum. However, three of those indices (H$_2$O-$J$, CH$_4$-$J$, and H$_2$O-$H$) are defined in spectral regions where unresolved absorption features (water) originate in \emph{both} the Earth's and the BD atmosphere. A division of the observed BD spectrum by that of a spectroscopic standard would tend to overcorrect for telluric absorption, even if both spectra were obtained under identical conditions, because within a resolution element the BD tends to be intrinsically fainter than the standard where the telluric absorption is stronger. Given the relatively large amount of precipitable water present during the observations, the use of these indices for spectral classification is hazardous. The two remaining indices are presented in Table~\ref{tbl-2} and yield a spectral type of T$2.5\pm0.5$.

\section{Discussion}\label{discussion}

SIMP 0136 was not identified as a BD by previous 2MASS searches because its NIR colors fall outside of the selection criteria used, e.g. $J-K_{\rm s} \le 0.3$ or $\ge 1.30$ \citep{Kirkpatrick1999}, $J-H\le0.3$ and $H-K_{\rm s}\le0$ \citep{Burgasser2003a}, or $J-K_{\rm s} \ge 1.0$ \citep{Cruz2003}. These selection criteria were defined to minimize the number of M dwarfs in the candidate sample. SIMP 0136 would have been easily detected by SDSS but this region of the sky was not surveyed. SIMP 0136 is the first T dwarf discovered through PM in the NIR; the only other T dwarf discovered through PM is $\varepsilon$~Indi~B \citep{Scholz2003}, a discovery made with optical data. The discovery of SIMP 0136 within a dataset covering only $\sim10\%$ of the sky lends credency to the claim that many relatively bright ($J<16$) early-T dwarfs have yet to be discovered. 

\begin{table}
\begin{center}
\caption{\label{tbl-2} Spectral indices}
\begin{tabular}{lcccc}
\hline\hline
Index & Numerator & Denominator & Value & Spectral type \\
\hline
CH$_4$-$H$ & $f_{1.635-1.675}$ & $f_{1.560-1.600}$ & 0.87 & T2\\
CH$_4$-$K$ & $f_{2.215-2.255}$ & $f_{2.080-2.120}$ & 0.42 & T3\\
\hline
\end{tabular}
\end{center}
\end{table}

The photometric distance of SIMP 0136 can be estimated using the distances and magnitudes of known T dwarfs of similar spectral types. Table~\ref{tbl-3} gives the NIR absolute magnitudes of the four T dwarfs in the T1-T4 range with distances known with an accuracy better than $\sim10\%$. Using their mean absolute magnitudes in $J$, $H$ and $K_s$, and assuming that SIMP 0136 has similar absolute magnitudes, we find photometric distances of $6.12\pm0.97$, $6.26\pm0.34$ and $6.61\pm0.41$~pc respectively. The weighted average photometric distance is $6.4\pm0.3$~pc.

\begin{table*}
\begin{center}
\caption{\label{tbl-3} Near-IR absolute magnitudes of nearby T dwarfs}
\begin{tabular}{lcr@{$\:\pm\:$}lccc}
\hline\hline
Name & Sp & \multicolumn{2}{c}{Distance} & \multicolumn{3}{c}{Absolute magnitude}\\
     &    & \multicolumn{2}{c}{(pc)}     & $J$ & $H$ & $K_s$ \\
\hline
SDSS0151+12          &T1   &  21.4&1.5$^{\ a}$   &$14.91\pm 0.20$&$13.95\pm0.19$&$13.53\pm 0.25$\\
SDSS1254-01          &T2   & 11.78&0.26$^{\ b}$  &$14.54\pm0.06$&$13.73\pm0.05$&$13.48\pm 0.07$\\
SDSS1750+17          &T3.5 &  27.6&3.5$^{\ a}$   &$14.14\pm0.29$&$13.75\pm0.30$&$13.27\pm 0.33$\\
$\varepsilon$~Indi~Ba &T1   & 3.626&0.009$^{\ c}$ &$14.49\pm0.02$&$13.71\pm0.02$&$13.55\pm0.02$\\
\hline
Mean&\multicolumn{3}{c}{}&$14.52\pm0.32$ & $13.79\pm0.11$&$13.46\pm0.13$\\
\hline
\end{tabular}
\tablenotetext{a}{Parallax from \citet{Vrba2004}.}
\tablenotetext{b}{Parallax from \citet{Dahn2002}.}
\tablenotetext{c}{Parallax from the Hipparcos catalog \citep{Perryman1997}.}
\end{center}
\end{table*}

Considering that $\sim$40\% of T dwarfs are resolved binaries \citep{Burgasser2006b}, it is crucial to test SIMP 0136 for binarity to ensure a proper interpretation of the observations and a legitimate comparison with models. Astrometric observations are underway to determine the parallax of SIMP 0136 using the NIR camera WIRCAM \citep{Puget2004} at the Canada-France-Hawaii Telescope. This camera features a field of view of $20^\prime\times20^\prime$ sampled with 0\farcs30 pixels. These observations have provided the best angular resolution observations for this object (0\farcs6 seeing) and no binarity or PSF elongation has been detected. Thus, SIMP 0136 does not appear to be a binary of mass ratio $\sim$1 and separation $\gtrsim$5 AU. However, observations at a higher resolution are needed to further constrain its multiplicity since most binary T dwarfs have separations smaller than this \citep{Burgasser2006b}. If SIMP 0136 is a binary, its photometric distance could be as large as $\sim$9 pc. The two other brightest early-T dwarfs, $\varepsilon$~Indi~Bab and SDSS0423-04 are both binaries \citep{Volk2003, McCaughrean2004, Burgasser2005}; thus, if SIMP 0136 is a single object, it would be the brightest isolated T dwarf, and as such it would become a benchmark object for the early-T spectral class, enabling very high-resolution spectroscopy, time-resolved spectroscopy and polarimetry observations. Its location near the celestial equator makes it all the more appealing as a benchmark object since it can be observed from all major observatories around the world.

\acknowledgments

The authors would like to thank Mike Shara for his leadership and financial support in enabling the CPAPIR observing campaign on the 1.5-m CTIO telescope. We are grateful to Adam Burgasser and the anonymous referee for their thoughtful comments on the manuscript. Our special thanks to Alberto Pasten and Claudio Aguilera for carrying out the SIMP observations, Philippe Vall\'ee for managing the SIMP data and Andr\'e-Nicolas Chen\'e for help with the spectroscopic observations. This work was supported in part through grants from the Natural Sciences and Engineering Research Council, Canada, and from Fonds Qu\'eb\'ecois de la Recherche sur la Nature et les Technologies. CPAPIR was built through a grant from the Canada Foundation for Innovation. This research has benefitted from the M, L, and T dwarf compendium housed at \mbox{DwarfArchives.org} and maintained by Chris Gelino, Davy Kirkpatrick, and Adam Burgasser. This publication makes use of data products from the Two Micron All Sky Survey, which is a joint project of the University of Massachusetts and the Infrared Processing and Analysis Center/California Institute of Technology, funded by the National Aeronautics and Space Administration and the National Science Foundation.

\end{document}